\PassOptionsToPackage{backref}{hyperref} 
\documentclass[preprints,article,accept,moreauthors,pdftex]{Definitions/mdpi}

\firstpage{1} 
\pubvolume{1}
\issuenum{1}
\articlenumber{0}
\pubyear{2022}
\copyrightyear{2022}
\externaleditor{{Academic Editor: Marc S. Paolella} 
}
\datereceived{10 June 2021} 
\dateaccepted{6 April 2022} 
\datepublished{8 April 2022} 
\hreflink{https://doi.org/} % If needed use \linebreak
\pdfoutput=1

\usepackage{color}
\usepackage{babel}
\usepackage{verbatim}
\usepackage{float}
\usepackage{enumitem}
\usepackage{amsmath}
\usepackage{amsthm}
\usepackage{graphicx}
\usepackage{setspace}
\usepackage{natbib}

\makeatletter

\providecommand{\tabularnewline}{\\}
\floatstyle{ruled}
\newfloat{algorithm}{tbp}{loa}
\providecommand{\algorithmname}{Algorithm}
\floatname{algorithm}{\protect\algorithmname}

\theoremstyle{plain}
\newtheorem{thm}{\protect\theoremname}
\theoremstyle{remark}

\@ifundefined{showcaptionsetup}{}{%
 \PassOptionsToPackage{caption=false}{subfig}}
\usepackage{subfig}
\makeatother

\providecommand{\remarkname}{Remark}
\providecommand{\theoremname}{Theorem}
\Title{Combining Predictions of Auto Insurance Claims}

\TitleCitation{Combining Predictions of Auto Insurance Claims}

\Author{{Chenglong Ye} 
 $^{1,}$*, Lin Zhang $^{2}$, Mingxuan Han $^{3}$, Yanjia Yu $^{4}$, Bingxin Zhao $^{4}$ and Yuhong Yang $^{4}$}

\AuthorNames{Chenglong Ye, Lin Zhang, Mingxuan Han, Yanjia Yu, Bingxin Zhao, Yuhong
Yang}

\AuthorCitation{Ye, Chenglong, Lin Zhang, Mingxuan Han, Yanjia Yu, Bingxin Zhao, and Yuhong Yang}
\address{%
$^{1}$ \quad {Dr. Bing Zhang} %MDPI: please check if the department name correct?. %CY: the department name is correct
 Department of Statistics, University of Kentucky, {317 Multidisciplinary Science Building,\linebreak 725 Rose St,} %MDPI: combined the address, please confirm. %CY: confirmed
 Lexington, KY 40536, USA  \\
$^{2}$ \quad First American Financial, {Santa Ana, CA 92707, USA;} %MDPI: please add city, post code and country. %CY: added
 {lizhang1@firstam.com} 
\\
$^{3}$ \quad School of Computing, University of Utah, {Salt Lake City, UT 84112, USA;} %MDPI: added city, post code and country, please confirm. %CY: confirmed
 u1209601@umail.Utah.edu\\
$^{4}$ \quad School of Statistics, University of Minnesota, {Minneapolis, MN 55455, USA;} %MDPI:  added city, post code and country, please confirm. %CY: confirmed
 {yuxxx748@umn.edu} %MDPI: This email is different from our system. Please confirm..%CY: confirmed
 (Y.Y.); zhao0600@umn.edu (B.Z.); {yyang@stat.umn.edu} %MDPI: This email is different from our system. Please confirm..%CY: confirmed
 (Y.Y.)}

\corres{Correspondence: chenglong.ye@uky.edu}

\abstract{This paper aims to better predict highly skewed auto insurance claims by combining candidate predictions. We analyze a version of the
Kangaroo Auto Insurance company data and study the effects of combining different
methods using five measures of prediction accuracy. The results show the following.
First, when there is an outstanding (in terms of Gini Index) prediction
among the candidates, the \textquotedblleft forecast
combination puzzle\textquotedblright{} phenomenon disappears. The
simple average method performs much worse than the more sophisticated
model combination methods, indicating that combining different methods could help us avoid performance degradation. Second, the choice of the prediction accuracy measure is crucial in defining the best candidate prediction for \textquotedblleft low
frequency and high severity\textquotedblright{} (LFHS) data. For example,
mean square error (MSE) does not distinguish well between model combination methods, as the values are close. Third, the performances of different model combination methods can differ drastically. We propose using a new model combination method, named ARM-Tweedie,
for such LFHS data; it benefits from an optimal rate of convergence and exhibits a desirable
performance in several measures for the Kangaroo data. Fourth, overall,
model combination methods improve the prediction accuracy
for auto insurance claim costs. In particular, Adaptive Regression by Mixing (ARM), ARM-Tweedie,
and constrained Linear Regression can improve forecast performance when there are only weak learners or when no dominant learner exists.}


\begin{document}
%%%%%%%%%%%%%%%%%%%%%%%%%%%%%%%%%%%%%%%%%%
\global\long\def\armt{\textrm{ARM\_T}}%

\section{Introduction}

\textls[-15]{The average countrywide insurance expenditure tends to rise from year
to year. Analyzing insurance data to predict future insurance claim
costs is of enormous interest to the insurance industry. In particular,
the accurate prediction of  claim cost is fundamental in determining policy
premiums, as it prevents potentially losing customers due to overcharging
and potential loss of profits due to undercharging.}

Non-life insurance data are distinct from common regression data due
to their \textquotedblleft low frequency and high severity\textquotedblright{} (LFHS)
characteristic---i.e., the distribution of the claim cost is highly
right-skewed and features a large point mass at zero. This paper focuses
on improving the prediction accuracy
for such insurance data by model combination/averaging.

Researchers have developed various methods for analyzing insurance data
in recent decades. \citet{bailey1960two} proposed
the minimum bias procedure as an insurance pricing technique for multi-dimensional
classification. However, the minimum bias procedure lacks a statistical
evaluation of the model. See \citet{feldblum2003minimum} for a detailed
overview of the minimum bias procedure and its extensions. In the late 1990s, the generalized linear models (GLM) framework \citep{10.2307/2344614}
was applied to model the insurance data; this is now the standard
method used in the insurance industry for modeling claim costs. \citet{jorgensen1994fitting}
proposed the classical compound Poisson--Gamma model, which assumes
the number of claims to follow a Poisson distribution and be independent
of the average claim cost that has a Gamma distribution. \citet{gschlossl2007spatial}
extended this approach and allowed dependency between the number
of claims and the claim size through a fully Bayesian approach. \citet{smyth2002fitting}
used double generalized linear models for the case where we only observe
the claim cost but not the frequency. Many authors have proposed methods
for insurance pricing using different frameworks other than GLM, including
quantile regression \citep{Heras:2018ii}, hierarchical modeling \citep{Frees:2008ga},
machine learning \citep{Kascelan:2015kb,Yang:2016bf}, the copula model
\citep{Czado:2012jz}, and the spatial model \citep{gschlossl2007spatial}.

Given the availability of many useful statistical models, empirical evidence has shown
that combining models, in general, is a robust and effective way to
improve predictive performance. Many works have improved the prediction accuracy by combining different models, which can be different types of models or same-type models with different tuning parameters. For
instance, \citet{wolpert1992stacked} proposed the use of Stacked Generalization
to take prediction results from the first-layer base learners as meta-features
to produce model-based combined forecasts in the second layer. A
gradient-boosting machine \citep{friedman2001greedy}, known as greedy
function approximation, suggests that using a weighted average of many weak
learners can produce an accurate prediction. \citet{yang2001adaptive}
proposed performing adaptive regression by mixing (ARM), a weighted average method that works well
for both parametric and nonparametric regression with an unknown error variance. \citet{hansen2012jackknife}
proposed Jackknife model averaging, which involves a linearly weighted average of
linear estimators searching for the optimal weight of each base regression
model%please check your meaning is retained
. \citet{doi:10.1080/01621459.2015.1115762} proposed the use of a weight
choice criterion for optimal estimation in generalized linear models.
We refer readers to \citet{JMLR:v15:wang14b} for a detailed literature
review on the theory and methodology of model combination.

However, in the specific context of insurance data, little research has been
carried out on combining predictions, except for \citet{ohlsson2008combining,doi:10.1002/sta4.180}. In particular, \citet{doi:10.1002/sta4.180} proposed a method to merge some levels of a categorical predictor in the model, which is a pre-step of applying model averaging. \citet{ohlsson2008combining} proposed to combine the generalized linear model and the credibility model, with  special focus on the car model classification problem for auto insurance. These two works are not directly related to combining predictions generated from different models for highly zero-inflated insurance data. To the best of our knowledge, no previous work has been done. Given the apparent importance of accurately predicting insurance claim costs, we propose a model combination method to capture such data characteristics.

Our paper focuses on improving the prediction accuracy of individual
\sloppy models/predictions by combining multiple predictions. We investigate
how different model combination methods perform under different measures
of prediction accuracy for LFHS data.
We propose a model combination method named ARM-Tweedie, assuming the
claim cost follows a Tweedie distribution. The Tweedie distribution family includes both continuous distributions (e.g., normal, Inverse Gaussian, gamma) and discrete distributions (e.g., Poisson, compound Poisson-gamma). In particular, we use the compound Poisson-gamma distribution in the Tweedie family (with the parameter $1<p<2$) since it allows a mixture of zeros and positive continuous numbers. It is a popular choice in the application of claim cost modeling.

The contributions of this paper are threefold. First, we design a novel model combination method for zero-inflated non-negative response data, where most current model combination methods fail to capture such a characteristic in theory. Second, we show that our method achieves the optimal rate of convergence offered by the candidates. From the risk-bound perspective, our method adapts to the optimal estimation of the mean function. Third, the conclusions of our analysis on a real-life data set provide both tools and guidance, especially to practitioners, on applying model combination methods to claim cost data for both adaptation and improvement. 

More specifically, we try to answer several interesting questions:
Do model combination methods improve over the best candidate prediction for
insurance data? Is the so-called ``forecast combination puzzle''  \citep{https://doi.org/10.1002/for.928,econometrics7030039}
still relevant when dealing with insurance data? Under different measures
of prediction accuracy, which model combination methods work the best? We
carry out a real-data analysis in this work. Thirteen analysts participated
by building models to predict the claim cost of each insurance
policy in a holdout data set. Based on their predictions, we apply different model combination
methods to obtain new predictions in the hope of achieving a higher prediction accuracy.
Different measures of prediction accuracy are considered due to the
existence of various constraints or preferences in practice. For example, a reasonable
prediction should identify the most costly customer and
provide the correct scale of the claim amount. Specifically, our paper
includes five measures: mean absolute error, root-mean-square error,
rebalanced root-mean-square error, the relative difference between
the total predicted cost and the actual total cost, and the Normalized
Gini index.

The remainder of this paper is organized as follows. We describe the general methodology in
Section \ref{sec:General-Methodology}, a data summary in Section
\ref{subsec:Data-summary-and}, a description of the project in Section \ref{subsec:Project-Description},
and the measures of performance in Section \ref{subsec:Measures-of-Performance}.
Section \ref{sec: information of pred} describes the performance
of the predictions provided by the analysts. The results of the model combination methods are given in Section \ref{sec:Model-Combining}, while
we introduce the proposed ARM-Tweedie method in Section \ref{subsec:ARM-Tweedie}.
We end our paper with a discussion in Section \ref{sec:Conclusion-and-Discussion}. The proof of the main theoretical result is included in {Appendix} %MDPI: added appendix a citation, please confirm. %CY: confirmed
 \ref{app}.

\section{\label{sec:General-Methodology}General Methodology}

In this section, we provide a detailed description of our research methodology.

\subsection{Data Summary\label{subsec:Data-summary-and}}

The Kangaroo Auto Insurance data \citep{de2008generalized} is based on one-year
vehicle insurance policies written in 2004 or 2005. The original data set is downloadable
from the R package ``insuranceData''. We added a random noise to each continuous
variable before releasing them to the data analysts. The perturbed data are available upon request. There are
67,856 policies and 10 variables in this dataset. The variable information
is presented in Table \ref{tab:Variable-information-of}.

\begin{table}[H]
%\begin{centering}
\caption{{Variable} %MDPI: removed the vertical line for tables, please confirm, please confirm all tables. %CY: confirmed
 description of the Kangaroo dataset. The variables in bold are directly related to the claim cost. The number in parentheses is the variance ratio (variance of perturbed ones to that of unperturbed ones) of each continuous variable. For the response {\it claimcost0}, noise is only added to the positive values.
\label{tab:Variable-information-of}}
\begin{tabularx}{\hsize}{@{}@{\extracolsep{\fill}}llll@{}}%{llll}
\toprule
\textbf{Variable} & \textbf{Description} & \textbf{Variable} & \textbf{Description}\\
\midrule
veh\_value  & (1.10) Vehicle value & gender & The gender of the driver\tabularnewline
veh\_body & The type of the vehicle body & area & Driver's area of residence\tabularnewline
veh\_age & The age group of the vehicle & agecat & Driver's age group\tabularnewline
{\textbf{claimcst0}} %MDPI: please check if the bold necessary.
  & (1.23) Total claim amount & exposure  & (0.91) The covered period\tabularnewline
\textbf{numclaims} & Number of claims & \textbf{clm} & Indicator if at least one claim\tabularnewline
\bottomrule
\end{tabularx}
%\par\end{centering}

\end{table}

\subsection{Project Description\label{subsec:Project-Description}}

\begin{figure}[H]
%\centering{}
\subfloat{\includegraphics[scale=0.35]{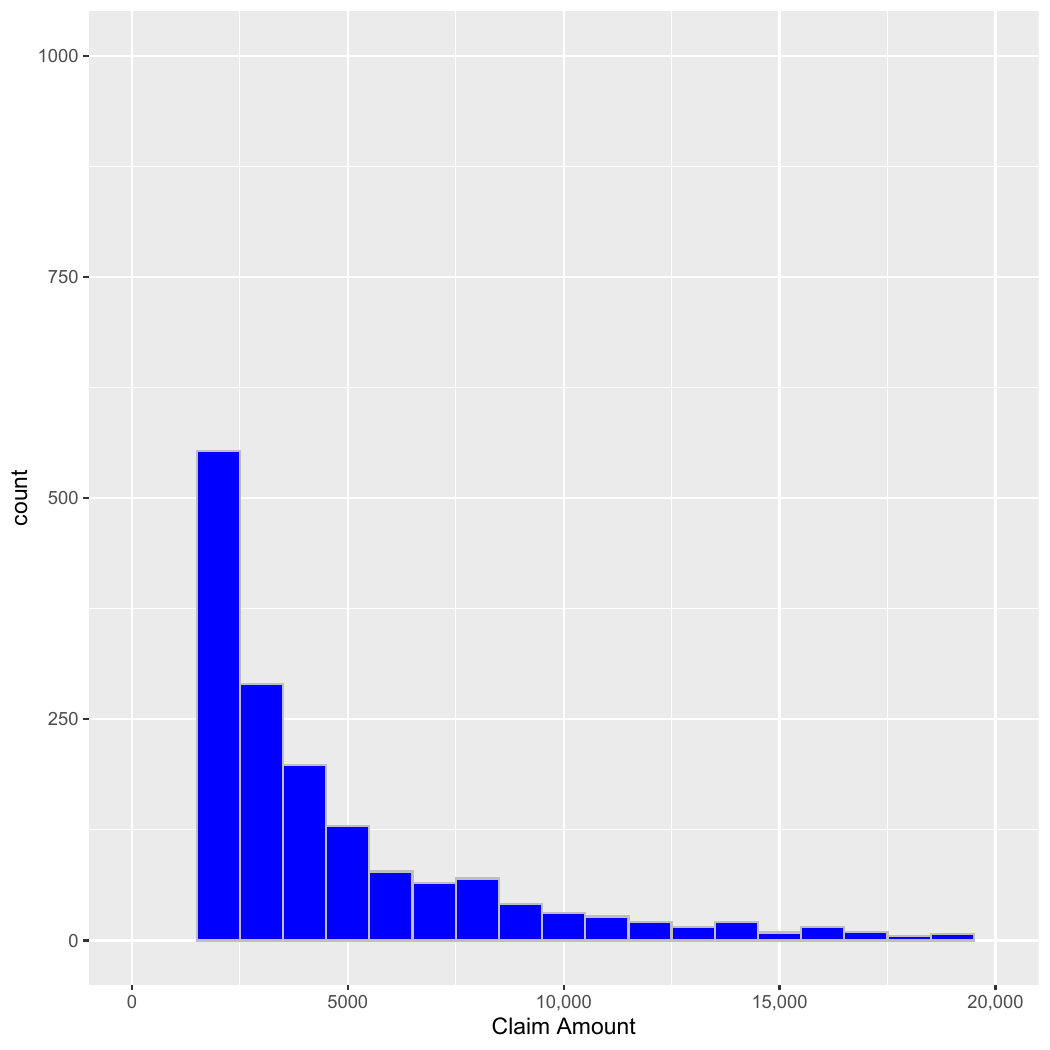}\label{fig:histclaim}} \subfloat{\includegraphics[scale=0.35]{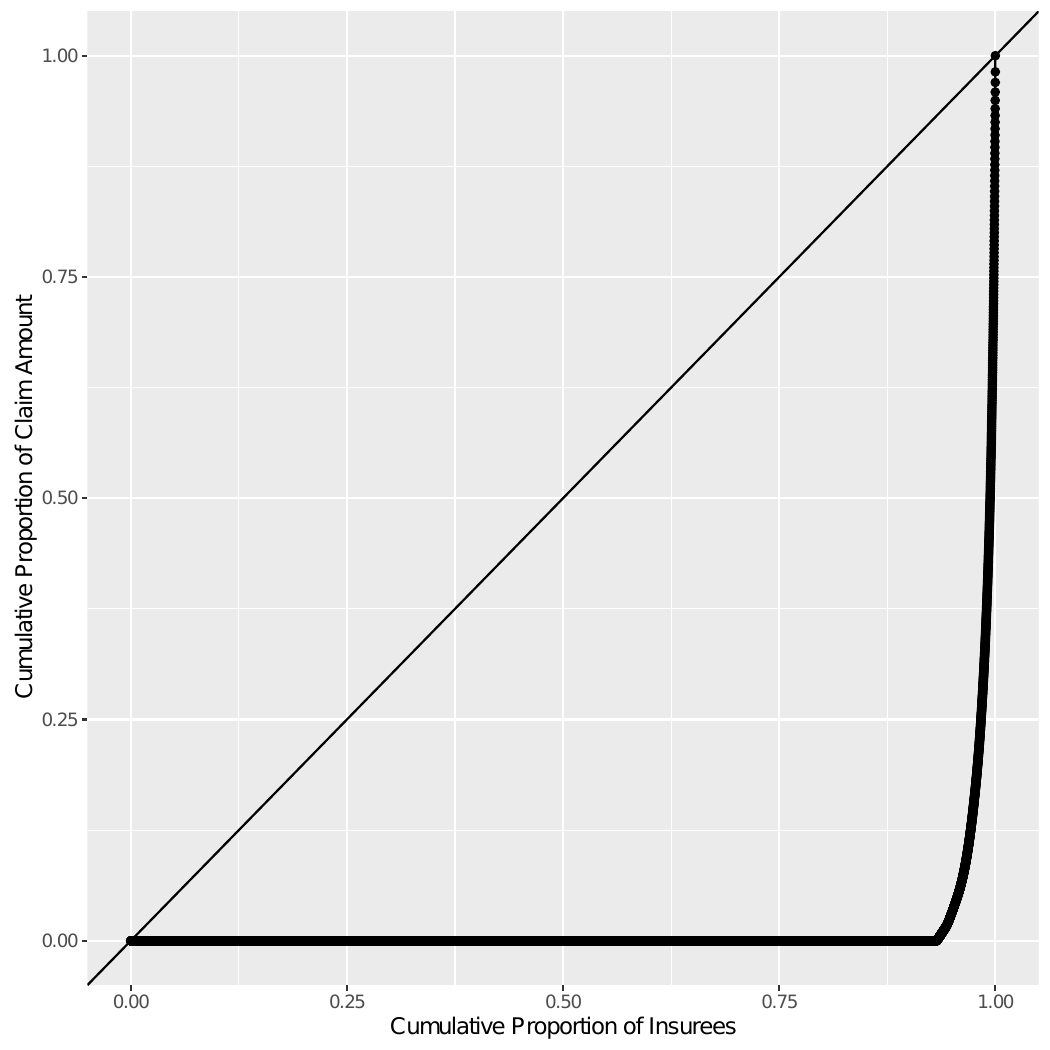}\label{fig:lorenz-1}} 
\caption{\label{fig:sd} {Data} %MDPI: more than 4-digit number need separated by comma, like chnage 10000 to 10,000, please confirm and revise.
 summary of the training set.  %MDPI: 2there is no subfigure number a and b, please confirm and revise.
 Left panel: histogram of the training set. There is a massive
spike at 0 with a frequency of 21,076, which is not plotted due to space
limitations. Right panel: Lorenz Curve of the training set.}
\end{figure}

We demonstrated the performance of the different model combination methods
for the Kangaroo data through the following procedure.
\begin{enumerate}
\item (Data Process) The dataset was split into 3 parts: 22,610 observations
for {T}%MDPI: is the bold necessary?.
raining, 22,629 observations for {V}alidation,
and 22,617 observations for {H}oldout;
\item (Prediction) Using only the training data, 13 analysts built
their models to predict the ``total amount of claims'' (\textit{claimcst0}).
We refer to these predictions made by the analysts as \textit{candidate predictions};
\item (Model Combination) We applied different model combination methods on the
13~candidate prediction models from step 2 and trained the model
averaging weights using a subset (5000 observations) of the validation
set.
\item (Evaluation) Finally, based on the holdout set, different predictive
performance measures were calculated for both the candidate predictions
and the combined predictions using model combination methods.
\end{enumerate}
\begin{Remark}
It is worth pointing out that 94\% of the claim costs have the value of zero (no
claims) in the training set. We present a histogram and a Lorenz curve
\citep{lorenz1905methods} (the cumulative proportion of the claim
amount against the cumulative proportion of insurers%please check your meaning is retained
) of the training
set in Figure \ref{fig:sd}. For the non-zero claims, the distribution
is right-skewed and heavy-tailed.
\end{Remark}

\begin{Remark}
In our description, it seems that not all the observations in the
validation set are used. Indeed, in step 2, we used the validation
set to evaluate the prediction accuracy of the candidate predictions.
Then, the analysts modified their models (they may also choose not to
modify their models). More accurately, the {\it candidate predictions} refer
to the predictions made after such modifications have taken place. For more details, see Section
\ref{subsec:Performance-of-combining}. In addition, the performances
of the 13 candidate predictions evaluated using the 5000 random observations
in step 3 are similar to the performances on the holdout set, verifying
the reasonability of using a sample of size 5000 to train the weights.
\end{Remark}
\vspace{-23pt}

\subsection{\label{subsec:Measures-of-Performance}Measure of Prediction Accuracy}

Let $n_e$ be the number of policies in the evaluation set. Denote $y_{i}$ and $\hat{y}_{i}$
as the claim cost and the predicted claim cost, respectively, for the
$i$-th policy. We consider the following five measures of the prediction
accuracy of $\{\hat{y}_{i}\}_{i=1}^{n_e}$.

\subsubsection*{Gini Index}

Gini index \citep{gini1912variabilita}, based on the ordered Lorenz
curve, is a well-accepted tool for evaluating the performance of auto
insurance claim predictions. There are many variants of the Gini
index. The one we utilize here is slightly different from those considered
in \citet{frees2014insurance}. 

For a sequence of numbers $\{s_{1},\dots,s_{n_e}\}$, let $\mbox{R}\left(s_{i}\right)\in\{1,\dots,n_e\}$
be the rank of $s_{i}$ in the sequence ($\mbox{R}\left(s_{i}\right)<\mbox{R}(s_{j})$
if $s_{i}<s_{j}$, given no ties exist. The tie-breaking method is
discussed in Remark \ref{rem:Unlike-other-measurements}). The
normalized Gini index is referred to as:
\begin{equation}
G=\dfrac{\sum_{i=1}^{n_e}y_{i}\mbox{R}\left(\hat{y}_{i}\right)\left/\sum_{i=1}^{n_e}y_{i}\right.-\sum_{i=1}^{n_e}\frac{n_e-i+1}{n_e}}{\sum_{i=1}^{n_e}y_{i}\mbox{R}\left(y_{i}\right)\left/\sum_{i=1}^{n_e}y_{i}\right.-\sum_{i=1}^{n_e}\frac{n_e-i+1}{n_e}}.\label{eq:1}
\end{equation}

\begin{Remark}
In (\ref{eq:1}), the Gini index depends on the predictions of $\{y_{i}\}_{i=1}^{n_e}$
only through their relative orders. Using some easy algebra, we obtain:
$\sum_{i=1}^{n_e}y_{i}\mbox{R}\left(y_{i}\right)\geq\sum_{i=1}^{n_e}y_{i}\mbox{R}\left(\hat{y}_{i}\right)$
and $\sum_{i=1}^{n_e}y_{i}\mbox{R}\left(y_{i}\right)+\sum_{i=1}^{n_e}y_{i}\mbox{R}\left(\hat{y}_{i}\right)\geq(n_e+1)\sum_{i=1}^{n_e}y_{i}$,
with \sloppy $\sum_{i=1}^{n_e}[y_{i}\mbox{R}\left(y_{i}\right)/\sum_{i=1}^{n_e}y_{i}]-\sum_{i=1}^{n_e}(n_e-i+1)/n_e>0$.
Therefore, we have $-1\leq G\leq1$, where the equality holds at $\mbox{R}\left(y_{i}\right)=\mbox{R}\left(\hat{y}_{i}\right)$
or $\mbox{R}\left(y_{i}\right)+\mbox{R}\left(\hat{y}_{i}\right)=n_e+1$,
respectively.
\end{Remark}

\begin{Remark}
\label{rem:Unlike-other-measurements}Unlike the other measures we consider,
a prediction with a larger Gini index (closer to 1) is favored. To break
the ties when calculating the order, we set $R(y_{i})>R(y_{j})$ if
$y_{i}=y_{j},\ i<j$.
\end{Remark}

\subsection{Root-Mean-Square Error (RMSE) and Mean Absolute Error (MAE)}

Root-mean-square error and mean absolute error are defined as $\sqrt{\frac{1}{n_e}\Sigma_{i=1}^{n_e}(y_{i}-\hat{y}_{i})^{2}}$
and $\frac{1}{n_e}\Sigma_{i=1}^{n_e}|y_{i}-\hat{y}_{i}|$, respectively.

Whatever the determination of the policy premiums is, the insurance
company needs to make profits and thus cares about the difference
between the total cost and the predicted total cost. Below, we consider
two measures of prediction accuracy that consider the overall scale
of the prediction.

\subsubsection{Rebalanced Root-Mean-Square Error (Re-RMSE)}

Let $\lambda=\frac{\Sigma y_{i}}{\Sigma\hat{y}_{i}}$ be the scale
parameter. Then, the rebalanced root-mean-square error is defined as
$\sqrt{\frac{1}{n_e}\Sigma_{i=1}^{n_e}(y_{i}-\lambda\hat{y}_{i})^{2}}$; this is the root-mean-square error %please check your meaning is retained
 of the scaled/rebalanced prediction
$\lambda\hat{y}_{i}$, whose total predicted cost is equal to the
actual total claim cost.

\subsubsection{SUM Error}

Here, we define (relative) SUM error as $\Sigma_{i=1}^{n_e}(\hat{y}_{i}-y_{i})/\Sigma_{i=1}^{n_e}y_{i}$,
which is the relative difference between the total predicted cost and the actual
total cost. SUM error is a way to measure the deviance of the total predicted
claim cost from the actual total claim cost. Note that a SUM error with a small absolute value is preferred.

\section{\label{sec: information of pred}Performances of the Candidate Predictions}

The 13 candidate predictions can be categorized into two types. One
type is based on distinct predictions of the number of claims (frequency)
and the claim cost (severity). This approach typically generates predictions
with values of zero. The other type directly predicts the claim
cost, typically producing many small non-zero-valued predictions. Four out of the 13 candidate predictions belong
to the first type (distinct predictions).

Table \ref{tab:after-feedback} shows the performances of the 13 candidate
predictions. We also provide in Table \ref{tab:inverse} the partial correlation matrix of the candidate predictions given the true value of the response. No prediction outperformed all its competitors in every measure of prediction accuracy. For instance, A5 has the largest/worst
RMSE among all the predictions, while its Gini index (0.95) is overwhelmingly
better than that of any other analyst (none of the values are more than 0.26). The
MAE values of the predictions are closely related to SUM. Since the
response $\{y_{i}\}_{i=1}^{n_e}$ contains too many zeros, a prediction
$\{\hat{y}_{i}\}_{i=1}^{n_e}$ will have a relatively small MAE if $\max\{y_{i}\}$
is small, such as A1 with a SUM around $-$1. For the SUM error, most predictions
have negative values, except A5. Specifically, the SUM errors of A1
and A2 almost reach $-$1. We checked the predictions
of A1 and A2 and found that all the predicted values were less than
10. In practice, it is unreasonable to use such small-scale values
as a final prediction of the claim cost, even with their acceptable
performance on MAE and Gini. Thus, we suggest the use of more than
one measure of prediction accuracy in this context.
\begin{table}[H]
\caption{\label{tab:after-feedback}{Performance} %MDPI: suggested move this table after closely its first citation in section 3, please confirm and revise.
 of the combined predictions. The highlighted values in each column indicate the
best model combination method for each scenario. We also provide the estimated
standard error of MAE, RMSE, and Re-RMSE to understand their reliability. The bold means the best performance among the 13 preditctions for each prediction measure. The N/A is because the combined prediction based on QR uses 0 as the
prediction for every observation, so the scale parameter in Re-RMSE
does not exist.}
%\begin{centering}
\begin{tabularx}{\hsize}{@{}@{\extracolsep{\fill}}llllll@{}}%{}
\toprule
\textbf{Predictions} & \textbf{MAE} & \textbf{RMSE} & \textbf{Re\_RMSE} & \textbf{Gini} & \textbf{SUM}\tabularnewline
\midrule
A1 & {\textbf{149.93}} %MDPI: is the bold necessary? %CY: Yes, the bold is necessary since the bold means the best performance among the 13 preditctions for each prediction measure.
 (7.49) & {1136.00 (65.71)} %MDPI: can add space? like change to 1136.00 (65.71)? please confirm whole table.
 & 1125.41 (65.57) & 0.1956 & $-$1.00\tabularnewline
A2 & 154.08 (7.48) & 1135.36 (65.72) & 1125.54 (65.45) & 0.2092 & $-$0.97\tabularnewline
A3 & 271.00 (7.26) & 1125.42 (65.55) & 1125.37 (65.51) & 0.1678 & \textbf{$-$0.05}\tabularnewline
A4 & 269.81 (7.26) & 1125.23 (65.46) & 1125.25 (65.41) & 0.1942 & \textbf{$-$0.05}\tabularnewline
A5 & 203.43 (8.35) & 1271.88 (57.76) & 1156.88 (58.40) & \textbf{0.9553} & 0.27\tabularnewline
A6 & 270.39 (7.27) & 1125.55 (65.20) & 1125.59 (65.12) & 0.1328 & $-$0.07\tabularnewline
A7 & 270.11 (7.26) & 1125.29 (65.33) & 1125.37 (65.27) & 0.2163 & \textbf{$-$0.05}\tabularnewline
A8 & 267.72 (7.26) & 1124.76 (65.46) & 1124.69 (65.40) & 0.2350 & $-$0.07\tabularnewline
A9 & 268.75 (7.26) & \textbf{1124.43} (65.44) & \textbf{1124.44} (65.38) & 0.2309 & \textbf{$-$0.05}\tabularnewline
A10 & 254.64 (7.30) & 1126.36 (65.59) & 1125.99 (65.45) & 0.1354 & $-$0.19\tabularnewline
A11 & 270.07 (7.26) & 1124.87 (65.37) & 1124.88 (65.31) & 0.2132 & \textbf{$-$0.05}\tabularnewline
A12 & 205.93 (7.38) & 1129.29 (65.78) & 1129.91 (65.36) & 0.1510 & $-$0.55\tabularnewline
A13 & 278.86 (7.24) & 1124.18 (65.27) & 1124.18 (65.28) & 0.2501 & 0.015\tabularnewline
\midrule 
\multicolumn{6}{c}{Scenario 1: Combining all predictions}\tabularnewline
\midrule 
SA & 228.21 (7.15) & 1099.16 (65.86) & 1092.37 (65.73) & 0.8707 & $-$0.216\tabularnewline
QR & \textbf{135.04} (7.09) & 1074.03 (65.88) & 1156.89 (58.39) & \textbf{0.9554} & $-$0.729\tabularnewline
ARM & 235.53 (6.91) & 1065.98 (65.50) & 1067.14 (65.57) & 0.9441 & \textbf{0.035}\tabularnewline
$\armt$ & 203.43 (8.35) & 1271.88 (57.76) & 1156.88 (58.40) & 0.9551 & 0.275\tabularnewline
GB & 135.42 (7.27) & 1101.76 (66.16) & \textbf{1002.43} (63.03) & 0.9307 & $-$0.859\tabularnewline
LR-C & 215.62 (6.89) & \textbf{1057.63} (63.68) & 1056.43 (64.05) & 0.9534 & 0.062\tabularnewline
\midrule 
\multicolumn{6}{c}{Scenario 2: Combining without A5}\tabularnewline
\midrule 
$\textrm{SA}_{(-5)}$ & 244.16 (7.30) & 1125.29 (65.40) & 1124.37 (65.60) & 0.2610 & $-$0.257\tabularnewline
QR$_{(-5)}$ & \textbf{149.90} (7.49) & 1136.01 (65.71) & N/A & $-$0.2519 & $-$1.000\tabularnewline
ARM$_{(-5)}$ & 272.38 (7.25) & 1123.75 (65.40) & 1123.69 (65.43) & 0.3127 & $-$0.032\tabularnewline
$\armt_{(-5)}$ & 270.19 (7.25) & 1123.86 (65.43) & 1123.75 (65.38) & 0.3166 & $-$0.052\tabularnewline
GB$_{(-5)}$ & \textbf{149.90} (7.49) & 1136.01 (65.42) & 1123.63 (65.71) & \textbf{0.3826} & $-$1.000\tabularnewline
LR-C$_{(-5)}$ & 273.70 (7.24) & \textbf{1123.36} (65.39) & \textbf{1123.32} (65.40) & 0.3300 & \textbf{$-$0.019}\tabularnewline
\midrule 
\multicolumn{6}{c}{Scenario 3: Combining Weak Learners (A1, A2, A3, A4, A6, A10, A12)}\tabularnewline
\midrule 
SA & 225.07 (7.34) & 1126.99 (65.69) & 1125.02 (65.43) & 0.2147 & $-$0.411\tabularnewline
QR & \textbf{149.90} (7.49) & 1136.01 (65.71) & N/A & $-$0.2519 & $-$1.000\tabularnewline
ARM & 269.73 (7.26) & \textbf{1124.97 (65.50)} & \textbf{1124.90 (65.45)} & \textbf{0.2236} & $-$0.059\tabularnewline
$\armt$ & 266.48 (7.27) & 1125.14 (65.54) & 1125.03 (65.48) & 0.2098 & $-$0.085\tabularnewline
GB & \textbf{149.90} (7.49) & 1136.01 (65.71) & 1125.67 (65.47) & 0.0347 & $-$1.000\tabularnewline
LR-C & 270.46 (7.26) & 1125.06 (65.53) & 1125.01 (65.49) & 0.2088 & \textbf{$-$0.052}\tabularnewline
\bottomrule

\end{tabularx}
%\par\end{centering}

\end{table}

\begin{table}[H]

\small

\caption{\label{tab:inverse}{The} %MDPI:  suggested move this table after closely its first citation in section 3, please confirm and revise.
 partial correlation matrix of the candidate predictions given the true value of \mbox{the response}.}
%\begin{centering}
\scalebox{1.012}{\begin{tabular}{cccccccccccccc}
\toprule
 &\textbf{ A1} & \textbf{A2} &\textbf{ A3} & \textbf{A4} &\textbf{ A5} & \textbf{A6} & \textbf{A7} & \textbf{A8} & \textbf{A9 }& \textbf{A10} & \textbf{A11} & \textbf{A12} & \textbf{A13}\tabularnewline
\midrule 
A1 & 1.00 & 0.81 & 0.91 & 0.78 & 0.06 & 0.19 & 0.66 & 0.75 & 0.76 & 0.81 & 0.70 & 0.55 & 0.03\tabularnewline
A2 & 0.81 & 1.00 & 0.61 & 0.96 & 0.06 & 0.55 & 0.80 & 0.86 & 0.85 & 0.59 & 0.89 & 0.50 & 0.05\tabularnewline
A3 & 0.91 & 0.61 & 1.00 & 0.57 & 0.06 & 0.02 & 0.49 & 0.60 & 0.57 & 0.78 & 0.54 & 0.47 & 0.03\tabularnewline
A4 & 0.78 & 0.96 & 0.57 & 1.00 & 0.05 & 0.55 & 0.78 & 0.83 & 0.78 & 0.56 & 0.85 & 0.49 & 0.05\tabularnewline
A5 & 0.06 & 0.06 & 0.06 & 0.05 & 1.00 & 0.03 & 0.06 & 0.06 & 0.07 & 0.04 & 0.06 & 0.05 & 0.00\tabularnewline
A6 & 0.19 & 0.55 & 0.02 & 0.55 & 0.03 & 1.00 & 0.82 & 0.72 & 0.54 & 0.06 & 0.74 & 0.25 & 0.12\tabularnewline
A7 & 0.66 & 0.80 & 0.49 & 0.78 & 0.06 & 0.82 & 1.00 & 0.96 & 0.81 & 0.47 & 0.89 & 0.49 & 0.12\tabularnewline
A8 & 0.75 & 0.86 & 0.60 & 0.83 & 0.06 & 0.72 & 0.96 & 1.00 & 0.85 & 0.54 & 0.93 & 0.55 & 0.12\tabularnewline
A9 & 0.76 & 0.85 & 0.57 & 0.78 & 0.07 & 0.54 & 0.81 & 0.85 & 1.00 & 0.50 & 0.81 & 0.46 & 0.05\tabularnewline
A10 & 0.81 & 0.59 & 0.78 & 0.56 & 0.04 & 0.06 & 0.47 & 0.54 & 0.50 & 1.00 & 0.49 & 0.42 & 0.02\tabularnewline
A11 & 0.70 & 0.89 & 0.54 & 0.85 & 0.06 & 0.74 & 0.89 & 0.93 & 0.81 & 0.49 & 1.00 & 0.51 & 0.16\tabularnewline
A12 & 0.55 & 0.50 & 0.47 & 0.49 & 0.05 & 0.25 & 0.49 & 0.55 & 0.46 & 0.42 & 0.51 & 1.00 & 0.01\tabularnewline
A13 & 0.03 & 0.05 & 0.03 & 0.05 & 0.00 & 0.12 & 0.12 & 0.12 & 0.05 & 0.02 & 0.16 & 0.01 & 1.00\tabularnewline
\bottomrule
\end{tabular}}
%\par\end{centering}
\end{table}

\section{\label{sec:Model-Combining}Model Combination}

Usually, model combination has two goals. Following the terms in \citep{yang2004combining,JMLR:v15:wang14b},
these are combining for {improvement} %MDPI: is tje bold necessary?.
 and combining
for {adaptation}. For improvement, we hope to combine the candidate
models to exceed the prediction performance of all the candidate models.
As for adaptation, it targets capturing the best model (usually unknown)
out of all the candidate models. In this paper, both goals are \mbox{of interest.}

Let $\mathbf{y}=\{y_{i}\}_{i\in Holdout}$ denote the response vector
for the holdout set. Denote $\mathbf{f}=(\mathbf{f}_{1},\dots,\mathbf{f}_{K})$
as the candidate prediction matrix, with each column representing a candidate prediction
to be combined for the holdout set. Let $\mathbf{f}_{c}=\sum_{k=1}^{K}\theta_k \mathbf{f}_k$ denote the
combined prediction. 

\subsection{Model Combination Methods}
\subsubsection{Some Existing Methods}

\paragraph{Simple Average (SA)}

The simple average method is the most basic procedure in model combination. We simply set $\theta_{k}\equiv\frac{1}{K},\ \forall k=1,\dots,K.$
In the literature, it is often reported that the simple average method has a better
or similar performance to that of other complicated methods; this is known
as the ``forecast combination puzzle'' \citep{https://doi.org/10.1002/for.928}. However, we are curious
about its performance in our case, where a dominant prediction exists
among the candidate predictions.

\paragraph{Linear Regression}

Treating the candidate predictions $\mathbf{f}=(\mathbf{f}_{1},\dots,\mathbf{f}_{K})$
as the regressors and $\mathbf{y}$ as the response, we fit a constrained
linear regression (LR-C): a linear regression of $\mathbf{y}$ on
$(\mathbf{f}_{1},\dots,\mathbf{f}_{K})$, with the constraint that
all the coefficients are non-negative and add up to 1. The estimated
coefficients become the corresponding weights for model combination.

We also tried the usual linear regression that allows negative coefficients. The performance (of the most interest, the Gini index is 0.93) is worse than linear regression with positive coefficients (Gini index being 0.95). Normalizing the coefficients by a positive number does not change the Gini index. So we decided not to present the usual linear regression and other methods, including quadratic optimization of the coefficients and linear regression with bounded coefficients. 

\paragraph*{Quantile Regression (QR) and Gradient Boosting (GB)}

We fit a quantile (median) regression model and a gradient boosting regression
model with candidate predictions as the features and $\mathbf{y}$
as the response. Then, the estimated coefficients will be the weights.
\begin{Remark}
The quantile regression predicts the median (when the quantile equals
0.5) rather than the mean of the response. In this case, we also use
the estimated coefficients as the weights in the combination. We consider
quantile regression because quantile regression does not require the
assumption of normality for error distribution and is robust to outliers.
\end{Remark}

\paragraph*{Adaptive Regression by Mixing (ARM)}

Adaptive regression by mixing, proposed by \citet{yang2001adaptive}, is a model combination method that involves data splitting
and cross-assessment.
\citet{yang2001adaptive} proves that the ARM weighting captures the optimal rate of convergence
among the candidate procedures for regression estimation. The advantage
is that under mild conditions, the resulting estimator is theoretically
shown to perform optimally in terms of rates of convergence without knowing
which candidate method works the best. Additionally, ARM typically works better
than AIC and BIC when the error variance is not small. In our application, we use the standard normal distribution for the noise distribution in ARM.

\subsubsection{ARM-Tweedie\label{subsec:ARM-Tweedie}}

In this subsection, we propose a model combination method for auto insurance
claim data. Consider a random variable $Y$ that belongs to the Tweedie
distribution family with a probability density function $f(y;\theta,\sigma^{2})=h(\sigma^{2},y)\exp\{\left(\theta y-b(\theta)\right)/\sigma^{2}\}$.
It is known that $E(Y)=b^{'}(\theta):=\mu$ and $Var(Y)=\sigma^{2}b^{''}(\theta)=\sigma^{2}\mu^{p}$
with the Tweedie power parameter $1<p<2$. Denote the above Tweedie
distribution as $f_{TW_{p}}(x;\mu,\sigma^{2})$ with the mean $\mu=(\frac{\theta}{\alpha-1})^{\alpha-1}$,
$\alpha=\frac{p-2}{p-1}$ and the dispersion parameter $\sigma^{2}$.
We assume that the data $\{y_i,x_i\}_{i=1}^n$ are generated from a Tweedie distribution 
\[
Y\sim f_{TW_{p}}(y;\mu=f(x),\sigma_{0}^{2}),
\]
where $\sigma_{0}^{2}$ is known. We assume that the distribution of the multivariate explanatory variable $x$ is $P(\cdot)$ and suppose that we have $f^{1},\dots,f^{K}$
as the candidate estimated functions for $f$.

We propose the following ARM-Tweedie algorithm (Algorithm \ref{alg:ARM-weighting-method-1}):

\begin{algorithm}[H]
\caption{{{}The ARM-Tweedie algorithm.}\label{alg:ARM-weighting-method-1}}
\begin{itemize}
\item Randomly and equally split the data into two subsamples $\mathbf{S}_{1}$
and $\mathbf{S}_{2}$.
\item For each $k$, implement the estimation procedure $f^{k}$ on $\mathbf{S}_{1}$
and obtain the estimated function $\hat{f}^{k}(x)$.
\item Compute the weight $w^{k}$: 
\[
w^{k}=\frac{\prod_{i\in \mathbf{S}_{2}}f_{TW_{p}}(y_{i};\hat{f}^{k}(x_i),\sigma_{0}^{2})}{\sum_{k\geq1}\prod_{i\in \mathbf{S}_{2}}f_{TW_{p}}(y_{i};\hat{f}^{k}(x_i),\sigma_{0}^{2})}.
\]
\item Repeat the above steps $L$ times and take the average as the final
weighting option: $w^{k}=\frac{1}{L}\sum_{l=1}^{L}w_{l}^{k}$.
\item Define $\hat{\delta}:=\sum_{k=1}^{K}w^{k}\hat{f}^{k}(x)$ as the combined
procedure.
\end{itemize}

\end{algorithm}

\begin{Remark}
In practice, we use the data $\mathbf{S}_{1}$ to obtain an estimator
of $\sigma_{0}^{2}$:
\[
\hat{\sigma}_{0}^{2}=\frac{\frac{1}{|\mathbf{S}_{1}|-1}\sum_{i\in\mathbf{S}_{1}}(y_{i}-\bar{y})^{2}}{(\frac{1}{|\mathbf{S}_{1}|}\sum_{i\in\mathbf{S}_{1}}y_{i})^{p}},
\]
where $\bar{y}=\sum_{i\in\mathbf{S}_{1}}y_{i}/|\mathbf{S}_{1}|$ and
$|\mathbf{S}_{1}|$ denotes the sample size of $\mathbf{S}_{1}$. Such an estimator is still plausible, although we allow nonparametric  $f$ as in $\mu=f(x)$. Given $\mu$, it is
still a parametric model in terms of the parameters of $\sigma^{2}$
and $p$. The estimator $\sigma_{0}^{2}$ only uses the
true value of the response $y_{i}$'s to estimate $\mu$, which is indeed a method of moment estimator regardless of the format of $f(x)$. 
\end{Remark}

\begin{Remark}
The value of $p$ is chosen as 1.5 in our specific data application. The Tweedie distribution
has two parameters: $\sigma^{2}$ and $p$. Given
$p$, the dispersion parameter $\sigma^{2}$ can be estimated by the
method of moment estimator as in Remark 6. The best
value of $p$ can be chosen by applying cross-validation on a set
of training data. In our data example, we found that the performance
of our method is quite stable against $p\in(1,2)$. So we set $p$
as the middle point of its range $(1,2)$ for simplicity.
\end{Remark}
\begin{Remark}
The procedures $f^{k}$'s are pre-determined by the researchers/practitioners.
For example, one can choose to directly apply a linear regression
to obtain the predictions for the claim cost. Since the prediction
for claim cost should be non-negative, we can set our final prediction
as zero if it is smaller than a cutoff and otherwise keep it unchanged.
Such a modeling procedure is considered $f^{1}$ and the estimated
predictions is denoted as $\hat{f}^{1}(x)$. Assume that a statistician
in an insurance company tries $K=10$ methods to predict the auto
insurance claim cost. It is also worth pointing out that our focus
is on the model combination stage. That is, we focus to further improve
the prediction accuracy by combining the 10 methods.
\end{Remark}

\begin{Assumption}
~
\begin{enumerate}
\item There exist two positive constants $B_{1},B_{2}$ such that $0<B_{1}<||f||_{\infty},||\hat{f}^{k}||_{\infty}<B_{2}$ at any $x$ for $k=1,\dots,K$, where  $||f||_{\infty}=\textrm{ess sup}|f|=\inf\{c\ge0:|f(X)|\le c\ \textrm{a.s.}\}$.
\item There exist two constants $\underline{\sigma}^{2},\overline{\sigma}^{2}$
such that $0<\underline{\sigma}^{2}\le\sigma_{0}^{2}\le\overline{\sigma}^{2}$.
\end{enumerate}
\end{Assumption}
{Let} %MDPI: added indentation, please confirm. %CY: Confirmed
 $||f||:=\sqrt{\int(f(x))^{2}P(dx)}$ for any function $f$. Let $n$ be the data sample size. Thus, we obtain the following theorem.
\begin{thm}
\label{thm:1}Suppose that Assumptions 1 and 2 hold. Based on a set of
estimation procedures $\{f^{k}\}_{k\ge1}$, the combined estimator
$\hat{\delta}$ constructed by the ARM-Tweedie algorithm has the following
risk bound:
\begin{eqnarray*}
E||f-\hat{\delta}||^{2}  & \leq & 4\left(B^2_{2}+\overline{\sigma}^{2}B^p_{2}\right)\inf_{k=1,\dots,K}\left\{ \frac{2}{n}\log K+C_{1}E||f-\hat{f}^{k}||^{2}\right\} ,
\end{eqnarray*}
where $C_{1}$ depends on $\text{\ensuremath{\sigma_{0}}}$ and $p$.
\end{thm}

The theorem indicates the adaptation of ARM-Tweedie for different
procedures.

\subsection{Performance of the Model Combination Methods \label{subsec:Performance-of-combining}}
We consider three scenarios based on the Gini index (of the most
interest) of the candidate predictions: (i) $\textrm{K}=13$---i.e., combining the 13 available
candidate predictions; (ii) $\textrm{K}=12$---i.e., there is no dominantly
better prediction (combining all the candidate predictions except
for A5); and (iii) $\textrm{K}=7$---i.e., all the candidate predictions
are weak (combining A1, A2, A3, A4, A6, A10, and A12, whose Gini index
is no greater than 0.2). The performance of some model combination methods varies drastically under these scenarios that are commonly encountered in practice.

Table \ref{tab:after-feedback} summarizes the performance of the
combined predictions under the five measures of prediction accuracy for each scenario.
Among all the model combination methods, ARM, ARM-Tweedie, and LR-C overall
perform well in both Gini and SUM. Note that Gini is only
related to the order of predictions, while SUM is more concerned with the
scale of the total cost of the claims. For RMSE and RE-RMSE, only small
differences are seen among the predictions, perhaps partly because of the large sample
size of the data. MAE is not suitable for measuring the prediction
performance alone. For example, in the table, QR$_{(-5)}$ (quantile
regression for combining all candidate predictions but A5) takes 0
as its prediction for every customer, giving no useful information.
However, the MAE of QR$_{(-5)}$ is the smallest. If one has to use a single
measure, Gini is recommended. Otherwise, we suggest the use of a combination
of at least two measures, including Gini.

From the perspective of a specific measure of prediction accuracy,
when there is a dominant candidate for prediction, such as A5 with
respect to the Gini index, it may be hard to achieve the goal of combining
for improvement. When there is no dominant candidate prediction, such
as under MAE, RMSE, Re-RMSE, and SUM in this paper, there is a better
chance of improving the performance through model combination. Specifically,
for MAE and RMSE, we have an approximately 10\% relative improvement
(from the best candidate prediction to the best combined prediction). For
Re-RMSE and SUM, the improvement is 25\% and 30\%, respectively. For
all the three scenarios, from the perspective of improving both Gini
and SUM, three methods (ARM, ARM-Tweedie, and LR-C) stand out from
all the model combination methods. It is also worth pointing out that GB or
QR can improve Gini or SUM, but not simultaneously. When there is
no dominant prediction, as in Scenarios 2 and 3, model combination methods
can improve the Gini index, even when there are only weak
learners.

The individual performance in Table \ref{tab:after-feedback} is the
second version of the models from the analysts. More specifically, when the analysts submitted their first prediction, the prediction performances evaluated on the validation set were provided.
Then, they modified their models (they were allowed not to modify them) and
submitted the second version of their predictions. Indeed, some analysts
changed their predictions significantly. For example, A8 has a negative
Gini index in the first version of predictions. However, the model combination
results are not very affected. This is because some candidate predictions
(more importantly, those with better predictive performance) show
little change after modification. Compared to the candidate predictions,
model combination methods are more stable than using a single method for predictive
modeling.

\section{\label{sec:Conclusion-and-Discussion}Conclusions and Discussion}

We start this section by answering the questions raised in the introduction.

\textbf{{Can model combination methods} %MDPI: is the bold necessary?. %CY: yes,, the three highlighted sentences are the key questions we try to answer in this paper.
 improve the results compared to the best individual prediction
when there is a dominant candidate prediction? }From our results, it is hard to achieve the goal of ``combining for improvement''
when there is a dominant candidate prediction. One reason for this may be that
these general model combination methods weaken the predictive power of the
dominant prediction. However, this does not exclude the possibility
that model combination methods unknown to us at this time can achieve a better predictive performance
than that of the best candidate. A follow-up question is: {when do model combination methods perform better than the best individual
prediction?} Based on our results, when all candidates are weak
or when no dominant candidate exists, model combination is a valuable
way to improve the \mbox{prediction performance.}

\textbf{{Does the ``forecast combination puzzle''} still exist in
our project for insurance data? } There are two possible scenarios where simple average outperforms other model combination methods. First, when all the candidates have the same level of bias, taking the average reduces the variability. Second, the biases among the candidates cancel each other out through the simple average method. However, our project concludes that the simple average
method does not provide competitive performance with that of other model combination methods. Specifically,
the Gini index of SA was the smallest and significantly worse than that of other model combination methods
in our results. The set of candidate predictions is of great importance
when considering the simple average method. When a dominant prediction exists
for a particular measure (the Gini index in our data analysis), simply averaging all the candidate predictions may lead to performance deterioration. In that case, we need to use a model combination method that adaptively learns
better from the data.

\textbf{{Under different measures of prediction} accuracy, which model combination
methods work the best? }When researchers and insurance companies are
concerned with different aspects of a prediction, their preferences
differ accordingly. For the criteria we considered, most combination
methods improve the performance of the best candidate prediction.
The measure is crucial in highly skewed zero-inflated data. We highly
recommend ``using at least two measures'' rather than just relying
on one single measure. For example, Gini is of the most interest when
evaluating the prediction of the claim cost. It only evaluates the rank of the predictions.
In the real world, the scale of the predicted claim cost is crucial
in determining the premium for a customer. Thus, if the Gini index
is large and the SUM is small in absolute value, the predictions do not need any scale
adjustment. Otherwise, a third measure such as RMSE should be considered
after adjusting the scales of the predictions. Based on our analysis,
we suggest not using MAE as a performance measure for predicting
the claim cost.

In our data analysis, the details of the generation of the 13 candidate models
are unknown. It is possible that two models were built using the same
model class but with different parameters, which may have led to a high
correlation between the two predictions. It would also be of interest
to study whether the details of the models will improve the performance
of the model combination methods. Additionally, it would be worth investigating a model combination method that assigns weights according to a specific performance measure (concerning the data type).
Another option for model combination is to combine all the subsets ($2^{13}$
candidate predictions), which may produce a higher variability or more potential \citep{JMLR:v15:wang14b} than combining
the 13 candidate predictions. However, this is more time-consuming.
This may even be computationally infeasible when the number of candidate predictions is large. One should consider the practical cost
when conducting model combination methods based on all the subsets. In addition, we may pay a much higher price in modeling variability when including all the subsets rather than the candidate predictions. In our project, combining
all the subsets led to a slightly better performance than combining the 13 candidate predictions
only in some cases; thus, we did not include the results in the table.

\authorcontributions{Conceptualization, Y.Y. (Yuhong Yang); methodology,  C.Y. and Y.Y. (Yuhong Yang); formal analysis, C.Y.,  L.Z., M.H., B.Z., and Y. Y. (Yanjia Yu); investigation, L.Z., M.H., B.Z. and Y. Y. (Yanjia Yu); data curation, C.Y.; writing---original draft preparation, C.Y.; writing---review and editing, C.Y. and Y.Y. (Yuhong Yang); supervision, Y.Y. (Yuhong Yang); project administration, C.Y. All authors have read and agreed to the published version of the manuscript.}

\funding{{This research received no external funding.} %MDPI: }Please add: ``This research received no external funding'' or ``This research was funded by NAME OF FUNDER grant number XXX.'' and  and ``The APC was funded by XXX''. Check carefully that the details given are accurate and use the standard spelling of funding agency names at \url{https://search.crossref.org/funding}, any errors may affect your future funding.
}

\institutionalreview{Not applicable.}

\informedconsent{Not applicable.} 

\dataavailability{The data presented in this study are available on request from the corresponding author.} 

\conflictsofinterest{The authors declare no conflict of interest.} 
\acknowledgments{We thank the anonymous reviewers and the Editor for their comments that improved this work. We also thank Dr. Zhuo Chen for his helpful comments.} 

%% Optional

%%%%%%%%%%%%%%%%%%%%%%%%%%%%%%%%%%%%%%%%%%
%% Only for journal Encyclopedia
%\entrylink{The Link to this entry published on the encyclopedia platform.}

%%%%%%%%%%%%%%%%%%%%%%%%%%%%%%%%%%%%%%%%%%
%% Optional
%\abbreviations{Abbreviations}{
%The following abbreviations are used in this manuscript:\\
%
%\noindent 
%\begin{tabular}{@{}ll}
%MDPI & Multidisciplinary Digital Publishing Institute\\
%DOAJ & Directory of open access journals\\
%TLA & Three letter acronym\\
%LD & Linear dichroism
%\end{tabular}}

%%%%%%%%%%%%%%%%%%%%%%%%%%%%%%%%%%%%%%%%%%
%% Optional
\appendixtitles{yes} % Leave argument "no" if all appendix headings stay EMPTY (then no dot is printed after "Appendix A"). If the appendix sections contain a heading then change the argument to "yes".
\appendixstart
\appendix
\section{Proof of Theorem \ref{thm:1}}\label{app}

\subsection{A More General Algorithm}

We first introduce a more general algorithm, of which ARM-Tweedie
is a special case. Then, in {Appendix} %MDPI: added `Appendix', please confrim. %CY: confirmed
 \ref{subsec:Proof} we prove that the
theorem holds for the general algorithm.

\begin{algorithm}[H]
\caption{{{}A more general ARM-Tweedie algorithm.}\label{alg:ARM-weighting-method-1-1}}
\begin{itemize}
\item Choose $N$, which is of the same order as $n$ and $1\leq N\leq n$.
Split the data into two subsamples $S^{(1)}=(x_{i},y_{i})_{i=1}^{N}$
and $S^{(2)}=(x_{i},y_{i})_{i=N+1}^{n}$.
\item For each $k$ and $1\leq l\leq N-1$, conduct the $k$-th estimation
procedure $f^{k}$ on the sample $\{S^{(2)},(x_{i},y_{i})_{i=1}^{l}\}$
and denote $\hat{f}_{l}^{k}$ as the estimated function.
\item Let $\pi^{k}$ be the initial weighting for the set of candidate estimation
procedures $\{f^{k}\}_{k\geq1}$. \\
Compute the weight $w_{i}^{k}$: 
\[
w_{i}^{k}=\begin{cases}
\pi^{k} & i=1,\\
\frac{\pi^{k}\prod_{l=1}^{i-1}f_{TW_{p}}(y_{l+1};\hat{f}_{i}^{k}(x_{l+1}),\hat{\sigma}_{0}^{2})}{\sum_{k}\pi^{k}\prod_{l=1}^{i-1}f_{TW_{p}}(y_{l+1};\hat{f}_{i}^{k}(x_{l+1}),\hat{\sigma}_{0}^{2})} & 2\leq i\leq N.
\end{cases}
\]
\item Define $\hat{\delta}:=\frac{1}{N}\sum_{i=1}^{N}\sum_{k=1}^{K}w_{i}^{k}\hat{f}_{i}^{k}(x)$
as the combined procedure.
\end{itemize}

\end{algorithm}

\subsection{{Proof for the more general Algorithm A1} %MDPI: can removed subsection, and chnage to \begin{proof} \end{proof}? please confirm and revise.
}\label{subsec:Proof}

Let $p_{f}(y|x)$ denote the conditional density of $y$ given $x$.
We have
\[
\log p_{f}(y|x)={\displaystyle \frac{1}{\sigma_{0}^{2}}(-\frac{(f(x))^{(1-p)}}{p-1}y+\frac{(f(x))^{2-p}}{p-2})+\log h(\sigma_{0},y),}
\]
where the corresponding distribution of $p_{f}(y|x)$ has mean $f(x)$ and variance
$\sigma_{0}^{2}f^{p}(x)$. Define 
\[
q_{i}(y|x)=\begin{cases}
 \sum_{k}\pi^{k}p_{\hat{f}_{1}^{k}}(y|x)  & i=1,\\
\frac{\sum_{k}\pi^{k}\left(\prod_{l=1}^{i-1}p_{\hat{f}_{l}^{k}}(y_{l+1}|x_{l+1})\right)p_{\hat{f}_{i}^{k}}(y|x)}{\sum_{k}\pi^{k}\left(\prod_{l=1}^{i-1}p_{\hat{f}_{l}^{k}}(y_{l+1}|x_{l+1})\right)}& 2\leq i\leq N.
\end{cases}
\]

{Then,} %MDPI: can add indentation?.
 the joint density of $(x,y)$ is $p_{f}(x,y)=p_{f}(y|x)\cdot f_{X}(x)$
and $q_{i}(x,y)=q_{i}(y|x)\cdot f_{X}(x)$ respectively. Let $\hat{p}(y|x):=\frac{1}{N}\sum_{i=1}^{N}q_{i}(y|x)$.
Notice that the corresponding mean to this density function is $\hat{\delta}$.
Then, we have 
\begin{flalign*}
 & \sum_{i=1}^{N}E\left[D(p_{f}(x,y)||q_{i}(x,y))\right]\\
= & \sum_{i=1}^{N}E\int p_{f}(y|x)\log\frac{p_{f}(y|x)}{q_{i}(y|x)}P(dx)\mu(dy)\\
= & \sum_{i=1}^{N}E\int p_{f}(y_{i+1}|x_{i+1})\log\frac{p_{f}(y_{i+1}|x_{i+1})}{q_{i}(y_{i+1}|x_{i+1})}P(dx_{i+1})\mu(dy_{i+1})\\
= & E\int[\prod_{i=1}^{N}p_{f}(y_{i+1}|x_{i+1})]\sum_{i=1}^{N}\log\frac{p_{f}(y_{i+1}|x_{i+1})}{q_{i}(y_{i+1}|x_{i+1})}P(dx_{2})\cdots P(dx_{N+1})\mu(dy_{2})\cdots\mu(dy_{N+1})\\
= & E\int[\prod_{i=1}^{N}p_{f}(y_{i+1}|x_{i+1})]\log\frac{\Pi_{i=1}^{N}p_{f}(y_{i+1}|x_{i+1})}{\Pi_{i=1}^{N}q_{i}(y_{i+1}|x_{i+1})}P(dx_{2})\cdots P(dx_{N+1})\mu(dy_{2})\cdots\mu(dy_{N+1})\\
\le & E\int[\prod_{i=1}^{N}p_{f}(y_{i+1}|x_{i+1})]\log\frac{\Pi_{i=1}^{N}p_{f}(y_{i+1}|x_{i+1})}{\pi^{k}\Pi_{i=1}^{N}p_{\hat{f}_{i}^{k}}(y_{i+1}|x_{i+1})}P(dx_{2})\dotsi P(dx_{N+1})\mu(dy_{2})\dotsi\mu(dy_{N+1})\\
= & E\int[\prod_{i=1}^{N}p_{f}(y_{i+1}|x_{i+1})]\log\frac{\Pi_{i=1}^{N}p_{f}(y_{i+1}|x_{i+1})}{\Pi_{i=1}^{N}p_{\hat{f}_{i}^{k}}(y_{i+1}|x_{i+1})}P(dx_{2})\dotsi P(dx_{N+1})\mu(dy_{2})\dotsi\mu(dy_{N+1})\\
 & +\log\frac{1}{\pi^{k}},
\end{flalign*}
where the fifth equality is due to the definition of $q_{i}(y|x)$
and the inequality holds for any $k\geq1$ because $q_{i}(y_{i+1}|x_{i+1})=\Sigma_{k}\pi^{k}p_{\hat{f}_{i}^{k}}(y_{i+1}|x_{i+1})$.
Also, we have
\begin{flalign*}
 & E\int[\prod_{i=1}^{N}p_{f}(y_{i+1}|x_{i+1})]\log\frac{\prod_{i=1}^{N}p_{f}(y_{i+1}|x_{i+1})}{\prod_{i=1}^{N}p_{\hat{f}_{i}^{k}}(y_{i+1}|x_{i+1})}P(dx_{2})\cdots P(dx_{N+1})\mu(dy_{2})\cdots\mu(dy_{N+1})\\
= & E\int[\prod_{i=1}^{N}p_{f}(y_{i+1}|x_{i+1})]\sum_{i=1}^{N}\log\frac{p_{f}(y_{i+1}|x_{i+1})}{p_{\hat{f}_{i}^{k}}(y_{i+1}|x_{i+1})}P(dx_{2})\cdots P(dx_{N+1})\mu(dy_{2})\cdots\mu(dy_{N+1})\\
= & \sum_{i=1}^{N}ED(p_{f}(x,y)||p_{\hat{f}_{i}^{k}}(x,y)),
\end{flalign*}
with
\begin{flalign*}
 & D(p_{f}(x,y)||p_{\hat{f}_{i}^{k}}(x,y))\\
= & \int\int p_{f}(y|x)\log\frac{p_{f}(y|x)}{p_{\hat{f}_{i}^{k}}(y|x)}\mu(dy)P(dx)\\
= & \sigma_{0}^{-2}\int\int p_{f}(y|x)(-\frac{f^{1-p}(x)-(\hat{f}_{i}^{k}(x))^{(1-p)}}{p-1}\cdot y+\frac{f^{2-p}(x)-(\hat{f}_{i}^{k}(x))^{2-p}}{p-2})\mu(dy)P(dx)\\
= & \sigma_{0}^{-2}\int(-\frac{f^{1-p}(x)-(\hat{f}_{i}^{k}(x))^{(1-p)}}{p-1}\cdot f(x)+\frac{f^{2-p}(x)-(\hat{f}_{i}^{k}(x))^{2-p}}{p-2})P(dx)\\
= & \frac{1}{\sigma_{0}^{2}(p-1)(2-p)}\int(2-p)(\hat{f}_{i}^{k}(x))^{(1-p)}f(x)-f^{2-p}(x)+(p-1)(\hat{f}_{i}^{k}(x))^{2-p}P(dx)\\
\leq & \frac{1}{\sigma_{0}^{2}(p-1)(2-p)}K\int((f(x))^{(2-p)/2}-[\hat{f}_{i}^{k}(x)]^{(2-p)/2})^{2}P(dx)\\
\le & \frac{1}{\sigma_{0}^{2}(p-1)(2-p)}2K(p-1)B_{1}^{-p}\int(f(x)-\hat{f}_{i}^{k}(x))^{2}P(dx)\\
:= & K_{1}||f-\hat{f}_{i}^{k}||^{2},
\end{flalign*}
where the first inequality holds for a large enough $K$ because $1<p<2$
and $0<B_{1}\le f,\hat{f}_{i}^{k}\le B_{2}$, and the second inequality
holds by taking Taylor's expansion of the function $f^{1-p/2}$ at
$\hat{f}_{i}^{k}$. Therefore, we have 
\[\sum_{i=1}^{N}ED(p_{f}(x,y)||q_{i}(x,y))\leq\log\frac{1}{\pi^{k}}+\sum_{i=1}^{N}ED(p_{f}||p_{\hat{f}_{i}^{k}})\leq\log\frac{1}{\pi^{k}}+K_{1}\sum_{i=1}^{N}E||f-\hat{f}_{i}^{k}||^{2}.
\]

{Because} K-L divergence $D(f||g)$ is convex on $g$, we have 
\[ED(p_{f}(x,y)||\hat{p}(y|x))\leq\frac{1}{N}\sum_{i=1}^{N}ED(p_{f}(x,y)||q_{i}(x,y)).\]
 
 {Thus}, 
\begin{equation}
ED(p_{f}(x,y)||\hat{p}(y|x))\leq\inf_{k}\left\{\frac{1}{N} \log\frac{1}{\pi^{k}}+\frac{K_{1}}{N}\sum_{i=1}^{N}E||f-\hat{f}_{i}^{k}||^{2}\right\} .\label{eq:1-1}
\end{equation}

{We} also have $\int(\sqrt{f(x)}-\sqrt{g(x)})^{2}\upsilon(dx)\leq\int f(x)\log\frac{f(x)}{g(x)}\upsilon(dx)$,
i.e., the Hellinger distance is bounded by the K-L divergence.

Next, we want to show that our estimator, which can be treated as
the expectation corresponding to the density $\hat{p}(y|x)$, has the desired upper bound as
stated in the theorem.\vspace{6pt}
\begin{flalign*}
 & \left(\int yp_{f}(y|x)\mu(dy)-\int y\hat{p}(y|x)\mu(dy)\right)^{2}\\
= & \left(\int y(p_{f}-\hat{p})\mu(dy)\right)^{2}\\
= & \left(\int y(\sqrt{p_{f}}+\sqrt{\hat{p}})(\sqrt{p_{f}}-\sqrt{\hat{p}})\mu(dy)\right)^{2}\\
\leq & \int y^{2}(\sqrt{p_{f}}+\sqrt{\hat{p}})^{2}\mu(dy)\cdot\int(\sqrt{p_{f}}-\sqrt{\hat{p}})^{2}\mu(dy)\\
\leq & 2\int y^{2}(p_{f}+\hat{p})\mu(dy)\cdot\int(\sqrt{p_{f}}-\sqrt{\hat{p}})^{2}\mu(dy)\\
= & 2[f^2(x)+\sigma_{0}^{2}f^p(x)+\int y^{2}\hat{p}(y|x)\mu(dy)]\cdot\int(\sqrt{p_{f}}-\sqrt{\hat{p}})^{2}\mu(dy)\\
\leq & 2[B^2_{2}+\overline{\sigma}^{2}B^p_{2}+\int y^{2}\hat{p}(y|x)\mu(dy)]\cdot D(p_{f}(y|x)||\hat{p}(y|x)))\\
\le & 4(B^2_{2}+\overline{\sigma}^{2}B^p_{2})\cdot D(p_{f}(y|x)||\hat{p}(y|x))),
\end{flalign*}
where the last inequality holds, since $\hat{p}$ is a convex combination
of $p_{\hat{f}_{i}^{k}}(y|x)$; by the boundedness assumption of
$\hat{f}_{i}^{k}$, we also have $\int y^{2}\hat{p}(y|x)\mu(dy)\leq B^2_{2}+\overline{\sigma}^{2}B^p_{2}$. Thus, 
\begin{flalign*}
 & E\int(f(x)-\hat{\delta})^{2}P(dx)\\
= & E\int(\int yp_{f}(y|x)\mu(dy)-\int y\hat{p}(y|x)\mu(dy))^{2}P(dx)\\
\leq & 4(B^2_{2}+\overline{\sigma}^{2}B^p_{2})\cdot E\int D(p_{f}(y|x)||\hat{p}(y|x))P(dx)\\
= & 4(B^2_{2}+\overline{\sigma}^{2}B^p_{2})\cdot ED(p_{f}(x,y)||\hat{p}(x,y))\\
\leq & 4(B^2_{2}+\overline{\sigma}^{2}B^p_{2})\inf_{k}\left\{ \frac{1}{N}\log\frac{1}{\pi^{k}}+\frac{K_{1}}{N}\sum_{i=1}^{N}E||f-\hat{f}_{i}^{k}||^{2}\right\},
\end{flalign*}
where the last inequality holds because of \eqref{eq:1-1}.

Recall $N$ is of the same order as $n$. The desired upper bound in the theorem follows.%MDPI: citation can't in equations, please move (\eqref{eq:1-1}----A1) to main text
%%%%%%%%%%%%%%%%%%%%%%%%%%%%%%%%%%%%%%%%%%
%\end{paracol}

\begin{adjustwidth}{-\extralength}{0cm}
%\centering %% If there is a figure in wide page, please release command \centering
\reftitle{References}

\end{adjustwidth}
%%%%%%%%%%%%%%%%%%%%%%%%%%%%%%%%%%%%%%%%%%

\end{document}